\documentclass[amsmath,amssymb,onecolumn,nofootinbib,notitlepage]{revtex4-1}
 \newcommand{\be}[1]{\begin{equation}\label{#1}}
 \newcommand{\ba}[1]{\begin{eqnarray}\label{#1}}
 \newcommand{\ep}[1]{\epsilon_{#1}}
  \newcommand{\tep}[1]{\tilde{\epsilon}_{#1}}
 
 \newcommand{\de}[1]{\delta_{#1}}
  \newcommand{\tde}[1]{\tilde{\delta}_{#1}}

 \newcommand{\rd}{{\rm d}}
 
 \newcommand{\re}{{\rm e}}
 \newcommand{\pa}[1]{\left(#1\right)}
 \newcommand{\paq}[1]{\left[#1\right]}

 \newcommand{\M}{{\rm M_{\rm P}}}
 
 \newcommand{\R}{{\mathcal R}}

 \def\ee{\end{equation}}
 \def\ea{\end{eqnarray}}

\usepackage{amsmath}

\usepackage{graphicx,latexsym}
\usepackage{graphicx,epsf}
\usepackage{color}
\usepackage{epsfig}
\usepackage{amsmath}
\usepackage[T1]{fontenc}
\usepackage[utf8]{inputenc}
\usepackage{tikz}
\usepackage{hyperref}
\usepackage{xcolor}
\usepackage[normalem]{ulem}
\begin{document}
\title{Reconstruction methods and the amplification of the inflationary spectrum}
\author{Leonardo Chataignier}
\email{leonardo.chataignier@unibo.it}
\affiliation{Dipartimento di Fisica e Astronomia, Universit\`{a} di Bologna,
via Irnerio 46, 40126 Bologna, Italy\\I.N.F.N., Sezione di Bologna, I.S. FLAG, viale B. Pichat 6/2, 40127 Bologna, Italy}
\author{Alexander~Yu.~Kamenshchik}
\email{kamenshchik@bo.infn.it}
\affiliation{Dipartimento di Fisica e Astronomia, Universit\`{a} di Bologna,
via Irnerio 46, 40126 Bologna, Italy\\I.N.F.N., Sezione di Bologna, I.S. FLAG, viale B. Pichat 6/2, 40127 Bologna, Italy}
\author{Alessandro Tronconi}
\email{tronconi@bo.infn.it}
\affiliation{Dipartimento di Fisica e Astronomia, Universit\`{a} di Bologna,
via Irnerio 46, 40126 Bologna, Italy\\I.N.F.N., Sezione di Bologna, I.S. FLAG, viale B. Pichat 6/2, 40127 Bologna, Italy}
\author{Giovanni Venturi}
\email{giovanni.venturi@bo.infn.it}
\affiliation{Dipartimento di Fisica e Astronomia, Universit\`{a} di Bologna,
via Irnerio 46, 40126 Bologna, Italy\\I.N.F.N., Sezione di Bologna, I.S. FLAG, viale B. Pichat 6/2, 40127 Bologna, Italy}

\begin{abstract}\vspace{-0.1cm}
We analyze the consequences of different evolutions of the Hubble parameter on the spectrum of scalar inflationary perturbations. The analysis is restricted to inflationary phases described by a transient evolution, when uncommon features arise in the inflationary spectra that may lead to an amplitude enhancement. We then discuss how the spectrum is, respectively, amplified or blue-tilted in the presence or absence of a growing solution of the Mukhanov-Sasaki equation. The cases of general relativity with a minimally coupled inflaton and that of induced gravity are considered explicitly. Finally, some remarks on constant roll inflation are discussed. 
\end{abstract}

\maketitle\vspace{-0.5cm}

\section{Introduction}
The possibility that a large amount of the dark matter (DM) content in our Universe is made of (primordial) black holes (PBHs) has been seriously considered in the past few years \cite{BHDM}. This idea seems compelling because it could improve our understanding of cosmological evolution and, in particular, of inflation \cite{inflation}. Moreover, the PBH hypothesis is also intriguing due to the increasing amount of direct and indirect observations of black holes (BHs) out of the astrophysical range, as well as the current lack of evidence for particle models of DM that go beyond the Standard Model of particle physics.

According to the present observational bounds \cite{carr}, it is possible that even the whole DM content of the Universe today is composed of PBHs originated from the collapse of matter overdensities in a certain wavelength interval of inflationary perturbations. In this scenario, the abundance of PBHs is related to the amplitude of the inflaton fluctuations, the enhancement of which must be by several orders of magnitude with respect to (w.r.t.) the amplitude probed by cosmic microwave background (CMB) radiation. Nonetheless, the microscopic physics that originate such a mechanism of amplification is still debated. For example, the amplification needed can be generated by a phase of ultraslow roll (USR) inflation in the presence of an inflection point of the inflaton potential \cite{USR}. This USR phase is the consequence of a transient period of inflatonary evolution, when slow-roll conditions are violated, and the inflaton then relaxes toward a de Sitter attractor. In contrast to the case of the fluctuations imprinted in the CMB, the perturbations \cite{pert} do not freeze at horizon exit in this case, as a growing solution of the Mukhanov-Sasaki (MS) equation is present, and it is responsible for the amplification of the modes. Other possibilities have been considered in the literature, such as an inflationary model able to generate a blue-tilted spectrum without the presence of the growing solution \cite{blue,CR2}.

In this article, these two mechanisms of amplification are considered. Instead of analyzing the possible consequences of different inflationary models obtained by varying  the form of the inflaton-gravity action, we shall here consider different evolutions of the Hubble parameter and correspondingly obtain the inflaton action. Within this approach, even if the inflaton potential cannot be exactly reconstructed, the features of the resulting spectra can still be calculated, and one may verify whether their amplitude is amplified. For simplicity, our starting point is the case of a minimally coupled inflaton, then some nonminimally coupled models are also investigated. Moreover, different techniques for the reconstruction are adopted.

The article is organized as follows. In Sec. II, we review the general formalism of the dynamics of the inflationary perturbations adopting a slightly unconventional formalism, and we derive the conditions for the existence of a growing solution in the MS equation or simply a blue-tilted spectrum in the absence of this solution. Furthermore, a useful relation between the odd and even slow-roll (SR) parameters in a certain hierarchy is obtained. This relation is valid for transient periods described by a certain time evolution, and it will be employed across the entire article. In Sec. III, different models are analyzed, and the procedure for reconstruction is illustrated. In Sec. IV, the application of the formalism to constant roll inflation is studied. Finally, the conclusions are drawn in Sec. V.

\section{Inflationary perturbations}
Let us first review the formalism of the inflationary perturbations. On adopting a slightly unconventional approach, we find the conditions that must hold in order to have an amplification of the inflationary spectrum either as the wave number $k$ grows or as time evolves. In a realistic inflationary scenario, wherein amplification starts at some given time, both mechanisms essentially lead to an enhancement of the shortest wavelength part of the spectrum ($k>k_{\rm CMB}$). The conditions are then expressed in a model-independent form, which is valid provided the SR parameters are ``constant'', and we use it in what follows to discuss different scenarios.

In general, after some manipulations, the MS equation takes the following form
\be{msGR}
v_k''+\pa{k^2-\frac{z''}{z}}v_k=0 \ , 
\ee
where the prime denotes the derivative w.r.t. conformal time $\eta$ and $z$ is a time-dependent function that depends on the specific model of inflation. For example, in the case of general relativity (GR) with a minimally coupled inflaton, one has $z=a\sqrt{\ep{1}}$, which leads to (see, e.g., \cite{GR})
\be{ddzozGR}
\frac{z''}{z}=a^2H^2\paq{2-\ep{1}+\ep{2}\pa{\frac{3}{2}+\frac{\ep{2}}{4}-\frac{\ep{1}}{2}+\frac{\ep{3}}{2}}}\equiv a^2H^2 f_{\rm MS}(\ep{i}) \ , 
\ee
with $\ep{1}=-\dot H/H^2$, $\ep{i+1}=\ep{i}^{-1}\rd \ep{i}/\rd N$ for $i>0$ and $N=\ln a$. The infinite set of $\ep{i}$'s form the so-called hierarchy of ``Hubble flow functions'' of SR parameters. It is important to note that, depending on the model of inflation, other hierarchies are commonly used, and they are associated with the evolution of different (homogeneous) degrees of freedom.

In general, one has
\be{ddzoz}
\frac{z''}{z}\equiv a^2 H^2 f_{\rm MS} \ , 
\ee
where $f_{\rm MS}$ is a dimensionless quantity that takes a different form depending on the inflationary model. It can then be expressed as a function of the SR parameters $\ep{i}$'s. 

It is now convenient to define the new independent variable $\xi=k/(aH)$, where $\rd \xi/\rd \eta=-aH(1-\ep{1})\xi<0$ during inflation. Due to
\be{deta}
\frac{\rd}{\rd \eta}=-aH(1-\ep{1})\xi\frac{\rd}{\rd \xi}
\ee
and 
\be{ddeta}
\frac{\rd^2}{\rd \eta^2}=a^2H^2(1-\ep{1})^2
\paq{\xi^2\frac{\rd^2}{\rd \xi^2}+\frac{\ep{1}\ep{2}}{(1-\ep{1})^2}\xi\frac{\rd}{\rd \xi}} \ ,
\ee
we are led to
\be{MSxi}
\paq{\xi^2\frac{\rd^2v_k}{\rd \xi^2}+\frac{\ep{1}\ep{2}}{(1-\ep{1})^2}\xi\frac{\rd v_k}{\rd \xi}}+\frac{\xi^2-f_{\rm MS}(\ep{i})}{(1-\ep{1})^2}v_k=0 \ .
\ee 
On rewriting the MS equation in terms of $\xi$ one eliminates its explicit dependence on $aH$.

In the regime where the SR parameters are constant and in the long wavelength limit ($\xi\rightarrow 0$), Eq. (\ref{MSxi}) can be algebraically solved and the features of the primordial spectra can be derived in a straightforward manner. Indeed, in this limit, the two independent solutions of Eq. (\ref{MSxi}) have the form $v_k=\xi^{\alpha}$, where $\alpha$ satisfies the algebraic equation 
\be{algMS}
\alpha^2+\paq{\frac{\ep{1}\ep{2}}{(1-\ep{1})^2}-1}\alpha-\frac{f_{\rm MS}(\ep{i})}{(1-\ep{1})^2}=0 \ ,
\ee 
with
\be{alpha12}
\alpha_{1,2}=\frac{-\paq{\frac{\ep{1}\ep{2}}{(1-\ep{1})^2}-1}\pm\sqrt{\paq{\frac{\ep{1}\ep{2}}{(1-\ep{1})^2}-1}^2+4\frac{f_{\rm MS}(\ep{i})}{(1-\ep{1})^2}}}{2} \ .
\ee
For instance, when $f_{\rm MS}$ is defined by Eq. (\ref{ddzozGR}), and in the pure de Sitter case ($\ep{i}=0$), we obtain
\be{alpha12dS}
\alpha_{1,2}=\frac{1\pm3}{2} \ .
\ee
For this case, the positive solution, $\alpha_1=2$, decreases in time, while the negative solution, $\alpha_2=-1$, increases, and it remains nontrivial in the limit $\xi\rightarrow 0$, which leads to
\be{vkds}
v_{k,{\rm dS}}\sim k^{-1/2} \pa{\frac{k}{aH}}^{-1} \ ,\ \mathcal{R}_{k,{\rm dS}}\sim k^{-3/2} \frac{aH}{z}= k^{-3/2}H \ ,
\ee
where $\mathcal{R}_k\equiv v_k/z$ is the curvature perturbation ($z=a$ in the de Sitter case), and the prefactor $k^{-1/2}$ is essentially fixed by the initial (Bunch-Davies) conditions. The quantity $\mathcal{R}_k$ is independent of time, and the spectral index can be straightforwardly computed to be
\be{nsm1def}
n_s-1=\frac{\rd\ln \Delta_s^2}{\rd \ln k} \ , 
\ee
with $\Delta_s^2\equiv |\mathcal{R}_{k,{\rm dS}}|^2k^3/(2\pi^2)$, which leads to the well-known de Sitter result $(n_s-1)_{\rm dS}=0$.

In the SR case ($|\ep{i}|\ll 1$), the SR parameters can be approximated by constants and the expressions (\ref{alpha12}) are still valid but must be expanded to first order for consistency. One then obtains
\be{solSR}
\alpha_{1,2}=\frac{1\pm\sqrt{9+12\ep{1}+6\ep{2}}}{2}\simeq\frac{1\pm\pa{3+2\ep{1}+\ep{2}}}{2} \ , 
\ee
which implies $(n_s-1)_{\rm SR}=-2\ep{1}-\ep{2}$.

We note that there is a caveat one must take into account for USR. In this case, one finds the same solutions for the $\alpha$'s as the de Sitter case, but the definition of the curvature perturbations is different, since $z_{\rm USR}\propto a\sqrt{\ep{1}}\rightarrow 0$. Then, the amplitude of primordial curvature perturbations depends on time and is amplified. In the USR case, the spectral index cannot be calculated analytically with the same procedure as illustrated for de Sitter and SR.

One can better illustrate the differences among the three cases just mentioned by solving the equation for $\mathcal{R}_k$, 
\be{eqRgr}
\R_k''+2\frac{z'}{z}\R_k+k^2 \R_k=0 \ .
\ee 
In terms of $\xi$, Eq. (\ref{eqRgr}) can be conveniently rewritten as 
\be{eqRxigr}
\xi^2\frac{\rd^2 \R_k}{\rd \xi^2}+\paq{\frac{\ep{1}\ep{2}-2\pa{1-\ep{1}}\frac{\rd \ln z}{\rd N}}{\pa{1-\ep{1}}^2}}\xi \frac{\rd \R_k}{\rd \xi}+\frac{\xi^2}{\pa{1-\ep{1}}^2}\R_k=0 \ .
\ee
In GR with a minimally coupled inflaton, we have $\rd \ln z/\rd N=1+\ep{2}/2$. Then, for constant SR parameters and in the long wavelength limit, the last term is negligible, and the equation admits a constant solution and a solution proportional to $\xi^\beta$, where
\be{betas}
\beta=\frac{3-4\ep{1}+\ep{2}+\ep{1}\pa{\ep{1}-2\ep{2}}}{\pa{1-\ep{1}}^2} \ .
\ee
If $\xi^\beta$ decreases in time, the constant solution dominates in the $\xi\rightarrow 0$ limit. This is what happens for de Sitter and SR. In contrast, if $\xi^\beta$ increases in time, it dominates in the $\xi\rightarrow 0$ limit. This is what occurs for USR leading to results that are very different from de Sitter and SR, namely, an amplitude of the spectrum that increases in time. The nonconstant solution is 
\be{Rknonc}
\R_k\propto \pa{\frac{k}{a H}}^\beta\sim \re^{-\beta\pa{1-\ep{1}}\,N} \ , 
\ee
and it increases or decreases depending on the sign of
\be{Phi}
\Phi\equiv\beta\pa{1-\ep{1}}=\frac{3-4\ep{1}+\ep{2}+\ep{1}\pa{\ep{1}-2\ep{2}}}{\pa{1-\ep{1}}} \ ,
\ee
increasing if $\Phi<0$ and decreasing if $\Phi>0$. Only in the latter case can the spectral index of the primordial spectrum be analytically calculated by using the definition (\ref{nsm1def}). For a general inflationary model, one finds
\be{DsGR}
\Delta_s^2\propto  k^{2+2\alpha_2}=k^{2-\paq{\frac{\ep{1}\ep{2}}{(1-\ep{1})^2}-1}-\sqrt{\paq{\frac{\ep{1}\ep{2}}{(1-\ep{1})^2}-1}^2+4\frac{f_{\rm MS}(\ep{i})}{(1-\ep{1})^2}}} \ ,
\ee
and
\be{gennsm1}
n_s-1=2-\paq{\frac{\ep{1}\ep{2}}{(1-\ep{1})^2}-1}-\sqrt{\paq{\frac{\ep{1}\ep{2}}{(1-\ep{1})^2}-1}^2+4\frac{f_{\rm MS}(\ep{i})}{(1-\ep{1})^2}}\ .
\ee

\subsection{Evolutions with ``constant'' SR parameters}
Let us now illustrate an important point. The results obtained above are exact when the SR parameters are constant. However, given the recursive definition of the SR parameters ($\ep{i+1}=\ep{i}^{-1}\rd \ep{i}/\rd N$), a constant set of $\ep{i}$'s corresponds to either $H={\rm const}$ and $\ep{i}=0$ (de Sitter case) or $\ep{1}={\rm const}$ and $\ep{i}=0$ for $i>1$ (power law inflation). It may thus seem redundant to present the general formalism for such a restricted range of applications. Nevertheless, we note that the above results can be applied to a wider set of problems. First, as we already mentioned, the general results for $\Phi$ and $n_s-1$ can be applied to the SR case, in which the expressions must be expanded to the first order for consistency, since the SR parameters are approximately constant when they are small. Furthermore, the large $a$ limit of some transient phase (such as the USR phase) leads to nontrivial sequences of ``constant'' SR parameters. In these cases, one obtains a hierarchy of, for example, $\ep{i}$'s with constant, nonzero SR parameters for either even or odd values of $i$, while the remaining SR parameters are zero. For instance, let $\ep{i}\stackrel{N\rightarrow \infty}{=}l_i+L_{i}(N)$ with $\lim_{N\rightarrow \infty}L_{i}(N)=0$. Then, due to their recursive definition, one obtains 
\be{derlim}
\ep{i}\ep{i+1}\equiv\frac{\rd\ep{i}}{\rd N}\stackrel{a\rightarrow \infty}{=}L_{i,N}(N) \ ,
\ee
which leads to $\lim_{N\rightarrow \infty}\ep{i+1}=0$, provided $\lim_{N\rightarrow \infty}L_{i,N}(N)=0$, and, in particular, 
\be{epi+1}
\ep{i+1}\stackrel{N\rightarrow \infty}{=}\frac{L_{i,N}(N)}{l_i+L_{i}(N)} \ .
\ee
Moreover, 
\be{epi+2}
\ep{i+2}\equiv\frac{\rd\ep{i+1}/\rd N}{\ep{i+1}}\stackrel{N\rightarrow \infty}{=}\frac{L_{i,NN}(N)}{L_{i,N}(N)}+\ep{i+1} \ .
\ee
Let us now suppose $L_i(N)\propto\re^{-\gamma N}\sim a^{-\gamma}$, with $\gamma>0$. In this case
\be{epi+2par}
\ep{i+2}\stackrel{N\rightarrow \infty}{=}-\gamma+\ep{i+1} \ ,
\ee
and the subsequent terms of the hierarchy take values equal to zero and $-\gamma$:
\be{limits}
\lim_{N\rightarrow \infty}\ep{i}=l_i,\;\lim_{N\rightarrow \infty}\ep{i+1+2n}=0,\; \lim_{N\rightarrow \infty}\ep{i+2n}=-\gamma \ .
\ee
Therefore, because of their definition, an infinite sequence of SR parameters may take alternate ``constant'' values in the large $a$ limit. This property is crucial in the analysis that follows, and it depends on the form of $L_{i}(N)$. Indeed, exponential forms lead to the result (\ref{limits}) but, in contrast, if $L_i\propto N^{-\gamma}$, then the sequence obtained is $\lim_{N\rightarrow \infty}\ep{j}=0$ for $j>i$.

It is also worthwhile to mention that similar results can be generalized to other hierarchies of SR parameters because they only depend on the recursive definition of the SR parameters [analogously to Eq. (\ref{derlim})] and on the form of $L_i$. For example, the same results can be extended to the hierarchy of ``scalar field flow functions'' that is defined by $\de{0}=\phi/\phi_0$ and $\de{i}\de{i+1}=\rd \de{i}/\rd N$. In general, the $\ep{i}$'s and the $\de{i}$'s are related through the homogeneous Friedmann and Klein-Gordon equations, and, in some scenarios, it is useful to use one or both hierarchies. 

\section{Model Reconstruction}
We are interested in reconstructing scalar field potentials that describe transient inflationary solutions, which are associated with varying SR parameters with a ``constant'' behavior in the future (and necessarily $\ep{1}<1$). Therefore, the results illustrated in the previous section can be adopted to study such models and to verify whether they can generate an amplification of the primordial spectrum. Finding the entire evolution of the scalar field is not necessary for this purpose, and we will only calculate the potential and the asymptotic behavior of the homogeneous quantities in terms of the corresponding SR parameters. The potentials that lead to an amplification can then be used to build an inflationary model that fits the CMB observations and which produces a large amount of DM in the form of PBHs at the end of inflation.
 
\subsection{GR with a minimally coupled inflaton}
To proceed with the reconstruction, let us first briefly review the homogeneous Einstein equation,
\be{fr0}
H^2=\frac{1}{3\M^2}\pa{\frac{1}{2}\dot\phi^2+V(\phi)} \ ,
\ee
\be{fr1}
\dot H=-\frac{\dot \phi^2}{2\M^2} \ ,
\ee
which leads to
\be{fr01}
\M^2H^2\pa{3-\ep{1}}=V \ .
\ee
This last equation can be used to reconstruct the potential. Equations (\ref{fr1}) and (\ref{fr01}) can be conveniently used for the reconstructions starting from some ansatz for $H=H(a)$. In this case, Eq. (\ref{fr1}) becomes
\be{fr1a}
\ep{1}=\frac{1}{2\M^2}\pa{\frac{\rd \phi}{\rd \ln a}}^2 \ ,
\ee
which can be integrated to obtain, when possible, $a=a(\phi)$.

Let us first consider the following evolution of the Hubble constant:
\be{Hans1}
H=H_0\pa{\alpha+\frac{A}{a^n}}^m \ , 
\ee
where $A,\alpha,n>0$. Similar to USR, the evolution described by Eq. (\ref{Hans1}) has a de Sitter attractor in the future, and, indeed, $H(a)$ is that of USR when $n=6$ and $m=1/2$. (It is interesting to note that this evolution represents a general solution in the model with a minimally coupled scalar field and a constant potential, or, in other words, in a universe driven by a mixture of two fluids: a cosmological constant and stiff matter. It is curious that $n=6$ and $m=1/4$ yield the general solution for the universe driven by the Chaplygin gas \cite{Chap}.) We also note that the transient is described by $A/a^n\sim \re^{-n N}$ and that a result similar to Eq. (\ref{limits}) is then expected. This is easily verified if we explicitly calculate the hierarchy of SR parameters:
\be{ep11}
\ep{1}=m\cdot n\frac{A}{\alpha a^n+A}=m\,\ep{3}=m\,\ep{5}=\dots\stackrel{a\rightarrow +\infty}{\longrightarrow}0 
\ee
and
\be{ep21}
\ep{2}=-n\frac{\alpha a^n}{\alpha a^n+A}=\ep{4}=\ep{6}=\dots\stackrel{a\rightarrow +\infty}{\longrightarrow}-n \ ,
\ee
where $a>\paq{(m\cdot n-1)A/\alpha}^{1/n}$ is necessary for inflation to occur. We can integrate and invert Eq. (\ref{fr1a}) to obtain
\be{phia1}
\exp\pa{\frac{\phi-\phi_0}{\M}\sqrt{\frac{n}{2m}}}=\frac{x+1}{x-1}\frac{x_0-1}{x_0+1} \ , 
\ee 
with $x\equiv A^{-1/2}\sqrt{\alpha a^n+A}$ and $x,x_0>1$. Notice that $\phi=\phi_0$ when $x=x_0$. Conversely, $\phi=\phi_{\infty}$, with
\be{phiinf1}
\phi_{\infty}\equiv\phi_0+\M\sqrt{\frac{2m}{n}}\ln B_0 \ ,
\ee
for $x\rightarrow \infty$. Equation (\ref{phia1}) can be solved for $x$, which yields
\be{x1}
x=\frac{\re^{\frac{\phi-\phi_0}{\M}\sqrt{\frac{n}{2m}}}+B_0}{\re^{\frac{\phi-\phi_0}{\M}\sqrt{\frac{n}{2m}}}-B_0} \ , 
\ee
where $B_0=(x_0-1)/(x_0+1)$, and the reconstructed potential is finally 
\be{pot1}
V=H_0^2\pa{\frac{\alpha x^2}{x^2-1}}^{2m}\pa{3-\frac{n\cdot m}{x^2}} \ .
\ee
For $n=6$ and $m=1/2$, one recovers a constant potential and the USR evolution, as expected. For other choices of the parameters $n$ and $m$, the expression for the potential in terms of $\phi$ is a complicated function with exponentials that need not be written here explicitly. However, this cumbersome expression is exact. Since the asymptotic behavior of the potential at $\phi \sim \phi_\infty$ determines the limiting values of the SR parameters, we simply give the form of $V$ around $\phi_\infty$, which is
\be{Vasy}
V\simeq 3H_0^2\alpha^{2m}\paq{1+\frac{n}{4}\pa{1-\frac{n}{2}}\pa{\frac{\phi-\phi_\infty}{\M}}^2} \ .
\ee
Finally, let us calculate the consequences of the background evolution given by Eq. (\ref{Hans1}) on the inflationary spectrum. The value of $\Phi$ is
\be{Phi1neg}
\Phi=\frac{3-4\ep{1}+\ep{2}+\ep{1}\pa{\ep{1}-2\ep{2}}}{\pa{1-\ep{1}}}\stackrel{a\rightarrow +\infty}{\longrightarrow}3-n \ , 
\ee
and for $n>3$ the curvature perturbations $\mathcal R_k$ are amplified, after their horizon exit, as time passes. In contrast, if $0<n<3$, from the constant solution for $\R_k$, one finds
\be{nsm11}
n_s-1=n>0 \ , 
\ee
which implies the amplitude is that of a blue-tilted spectrum, which grows as the wave number $k$ increases. We conclude that for GR with a minimally coupled inflaton, the inflationary evolution described by Eq. (\ref{Hans1}), with a transient phase and a de Sitter attractor in the future, leads to an inflationary enhancement. The corresponding inflaton dynamics is driven by the potential (\ref{pot1}), and similar behaviors can be obtained from potentials of the form (\ref{Vasy}) with the field close to $\phi_\infty$.

\subsection{Power law solutions}
In this section, we generalize the results obtained from Eq. (\ref{Hans1}) and study the transient phase with a power law inflation attractor. For this case, in contrast to de Sitter, it is only possible to reconstruct the inflaton potential exactly for particular choices of the parameters. Close to the attractor, an approximate reconstruction can always be obtained, and that is enough for the purposes of model building. The amplification of the primordial spectrum can still be studied in full generality, as it depends on the asymptotic values of the SR parameters, which can be calculated exactly. In this case, and in the large $a$ limit, one obtains
\be{PLlim}
\ep{1}\rightarrow {\rm const}+L(a) \ , 
\ee  
with $L(a)\rightarrow 0$. In analogy to the previous case, we consider
\be{ep1PL}
\ep{1}=\pa{\beta+\frac{B}{a^n}}^m\rightarrow \beta^m \ , 
\ee
with $\beta,B,n>0$ and $\beta^m<1$ (so as to have acceleration close to the attractor). Notice that, when $\beta=0$, one finds a transient phase with a de Sitter attractor, but $\ep{1}$ in Eq. (\ref{ep1PL}) is different from that in the set (\ref{ep11}). This case is expected to generate a hierarchy of the form (\ref{limits}) in the large $a$ limit. Indeed, the ansatz (\ref{ep1PL}) leads to the following hierarchy of SR parameters:
\be{epeven2}
\ep{2}=-m\ep{4}=-m\ep{6}=\dots=-\frac{n\,m\,B}{B+\beta\,a^n}\rightarrow 0 
\ee
and
\be{epodd2}
\ep{3}=\ep{5}=\dots=-\frac{n\,\beta a^n}{B+\beta\,a^n}\rightarrow -n \ ,
\ee
where, in contrast to the de Sitter case examined in the previous section, now the even SR parameters tend to zero.

By proceeding with reconstruction and integrating Eqs. (\ref{ep1PL}) and (\ref{fr1a}), one finds, respectively,
\be{H2}
H=H_0\exp\paq{-\frac{\pa{\beta +\frac{B}{a^n}}^{1+m}}{(1+m)n\beta}\!\!\!{\phantom B}_2F_1\pa{1,1+m,2+m,1+\frac{B}{\beta a^n}}} 
\ee
and
\be{phi2}
\phi-\phi_0=f(a)-f(a_0) \ ,
\ee
where
\be{fa}
f(a)=\frac{2\sqrt{2}\M\pa{\beta+\frac{B}{a^n}}^{\frac{2+m}{2}}\!\!\!{\phantom B}_2F_1\pa{1,1+\frac{m}{2},2+\frac{m}{2},1+\frac{B}{\beta a^n}}}{(2+m)n\beta} \ .
\ee
In this case, the exact reconstruction of the potential is rather complicated unless one adopts simplifying assumptions. For example, let $m=-1$ and $0<\beta<1$. Then,
\be{H2m1}
H=\frac{H_0}{\paq{n\pa{B+\beta a^n}}^{\frac{1}{n\beta}}} 
\ee
and
\be{phi2m1}
\phi-\phi_0=\M\frac{\ln\pa{\frac{1+\sqrt{\frac{\ep1(a)}{\beta}}}{1-\sqrt{\frac{\ep1(a)}{\beta}}}\frac{1-\sqrt{\frac{\ep1(a_0)}{\beta}}}{1+\sqrt{\frac{\ep1(a_0)}{\beta}}}}}{n\sqrt{\beta}} \ .
\ee
In the $a\rightarrow \infty$ limit, one obtains
\be{phiinf2}
\phi_\infty=\phi_0+\M\frac{\ln\pa{\frac{\beta+1}{\beta-1}A_0}}{n\sqrt{\beta}} \ ,
\ee
with $A_0\equiv \pa{1-\sqrt{\ep1(a_0)/\beta}}/\pa{1+\sqrt{\ep1(a_0)/\beta}}$. The relation (\ref{phi2m1}) can be inverted to obtain $a^n=a^n(\phi)$:
\be{aphi2}
a^n=a_0^n\frac{\paq{\pa{1-\sqrt{\frac{\ep1(a_0)}{\beta}}}+\pa{1+\sqrt{\frac{\ep1(a_0)}{\beta}}}\re^{n\sqrt{\beta}(\phi-\phi_0)/\M}}^2}{4 \re^{n\sqrt{\beta}(\phi-\phi_0)/\M}} \ , 
\ee
and finally the potential can be reconstructed, provided we substitute Eq. (\ref{aphi2}) into Eq. (\ref{fr01}). In terms of $a^n$, it then takes the following form:
\be{pot2}
V=\frac{H_0^2}{\paq{n\pa{B+\beta a^n}}^{\frac{2}{n\beta}}}\paq{3-\pa{\beta+\frac{B}{a^n}}^m} \ .
\ee
The expression in terms of $\phi$ is cumbersome and it will not be needed. It is also worthwhile to note that such a potential depends on the homogeneous inflaton through the exponential function $\exp\pa{n\sqrt{\beta}\phi/\M}$. This functional dependence is expected as it is the generalization of the standard power law inflation potential, which contains only one exponential function of the inflaton. Moreover, various approximate reconstruction methods can be used to obtain the shape of the potential close to the attractor, but we omit this discussion here. Whereas the exact reconstruction can be obtained for certain values of the parameters, the behavior of the resulting inflationary spectra can be calculated exactly from Eqs. (\ref{epeven2}) and (\ref{epodd2}). For generic values of $m$, one may compute
\be{Phi2}
\Phi=\frac{\beta^{2m}-4\beta^m+3}{\pa{1-\beta^m}}=3-\beta^m>0 \ .
\ee
This shows the absence of the growing solution for Eq. (\ref{eqRgr}). The spectral index is then simply given by 
\be{nsm12}
n_s-1=-\frac{2\beta^m}{1-\beta^m}<0 \ .
\ee
This is the same result as the one obtained for the power law attractor solution. In contrast to the de Sitter case, the resulting primordial spectrum, if evaluated on the trajectory that approaches the attractor (and close to it), coincides with the spectrum calculated on the attractor itself, and no amplification or peculiar features emerge. It is also noteworthy that this result is not restricted to the evolution given by Eq. (\ref{ep1PL}), as it only depends on the limits (\ref{epeven2}) and (\ref{epodd2}), which are not particular to Eq. (\ref{ep1PL}). For instance, starting from the ansatz
\be{PLans2}
H(a)=H_0\pa{\frac{a_0}{a}}^{\beta^m}\pa{1+\frac{A_0}{a^n}}^m \ , 
\ee
where $n,A_0>0$, the resulting spectra are the same.

The observed absence of amplification in the cases of power law inflation considered here is relevant because power law is the exactly solvable inflationary model that is most akin to SR. One may then conjecture that similar results (and, in particular, the lack of amplification) hold for SR inflation when the inflaton approaches the attractor solution, and the enhancement is a peculiarity of de Sitter. 

\subsection{Nonminimally coupled inflaton}
Let us now consider the different scenario of a nonminimally coupled inflaton.  In order to perform the reconstruction, we first review the basic homogeneous equations for this model: 
\be{fr0NM}
H^2=\frac{1}{3F(\phi)}\pa{\frac{\dot \phi^2}{2}+V-3HF_{,\phi}\dot \phi} 
\ee 
and
\be{fr1NM}
\dot H=-\frac{1}{2F(\phi)}\paq{\pa{1+F_{,\phi\phi}}\dot\phi^2+F_{,\phi}\pa{\ddot \phi-H\dot \phi}} \ , 
\ee
where $F(\phi)$ represents a general nonminimal coupling and $F=\M$ reproduces the minimally coupled case. In contrast to the previous cases, the homogeneous equations and the reconstruction procedure now become more involved. Then, for simplicity, we shall henceforth limit our study to the induced gravity (IG) case, where $F(\phi)=\xi \phi^2$ \cite{IG0,IG}. This simplifying choice is also justified by the fact that both Higgs inflation and Starobinsky inflation (in the Einstein frame) occur in a regime that is very close to pure IG.

Reconstructing the inflaton potential for a given $H(a)$ is not as straightforward as for GR with a minimally coupled inflaton, and we found exact potentials only for certain values of the parameters and for the de Sitter attractor case [cf. Eq. (\ref{Hans1})]. Nevertheless, we can still predict the shape of the inflationary spectra or, at least, the possibility of an amplification in the large $a$ limit.

\subsection{De Sitter limit}
Let us consider $H(a)$ given by Eq. (\ref{Hans1}). In IG, the following exact relations hold between some SR parameters:
\be{rel1IG}
\ep{1}=\frac{\de{1}}{1+\de{1}}\pa{\frac{\de{1}}{2\xi}+2\de{1}+\de{2}-1} \ ,
\ee
\be{rel2IG}
\ep{1}=\frac{1}{2\xi(1+6\xi)}\paq{(1+2\xi)\de{1}^2-8\xi\de{1}-6\xi^2\pa{1+2\de{1}-\frac{\de{1}^2}{6\xi}}\pa{\frac{\rd \ln V}{\rd \ln \phi}-4}} \ .
\ee
Before discussing the reconstruction of the inflaton potential, we must first calculate the asymptotic values of the SR parameters, which are pivotal in the analysis of the amplification of the spectrum. From Eq. (\ref{fr0NM}), the potential can then be obtained as
\be{VrecIG}
V=3\xi\phi^2 H^2\pa{1+2\de{1}-\frac{1}{6\xi}\de{1}^2} \ , 
\ee 
provided $H=H(\phi)$ and $\de{1}=\de{1}(\phi)$ are known [indeed, Eq. (\ref{VrecIG}) is the IG counterpart of Eq. (\ref{fr01}) in GR].

Let us first calculate the SR parameters in the large $a$ limit. Since $\lim_{a\rightarrow \infty}\ep{1}=0$, one has either $\lim_{a\rightarrow \infty}\de{1}=0$ and $\lim_{a\rightarrow \infty}\de{2}\neq 0$, or $\lim_{a\rightarrow \infty}\de{2}=0$ and $\lim_{a\rightarrow \infty}\de{1}\neq 0$ but satisfying the relation
\be{d1inf}
\de{1,\infty}=\frac{2\xi}{1+4\xi} \ .
\ee
These results follow from the functional dependence of $H(a)$ on $a$ inherited by $\ep{1}$ and $\de{i}$'s and by the general result given in Eq. (\ref{limits}), which is applied here to the SR hierarchy $\de{i}$. Notice that, in contrast to GR, two different de Sitter trajectories are present in IG, and they are associated with two different evolutions of the inflaton field. Using Eq. (\ref{rel2IG}) in the same limit for $a$, one obtains that the potential, on the attractor, must satisfy
\be{fr0l1}
\frac{\rd \ln V_\infty}{\rd \ln \phi}-4=0\Rightarrow V_\infty\propto \phi^4
\ee
in the former case and
\be{fr0l2}
\frac{\rd \ln V_\infty}{\rd \ln \phi}-4=0\Rightarrow V_\infty\propto \phi^2
\ee
in the latter case.

We can now proceed to evaluate the full hierarchy of $\de{i}$'s. Starting from Eq. (\ref{rel1IG}) and differentiating, we find
\be{ep2IG}
\ep{2}=\frac{\de{2}\paq{(1+4\xi)\de{1}^2+2\xi\pa{\de{2}+\de{3}-1}+2\de{1}\pa{1+4\xi+\xi\de{3}}}}{(1+\de{1})\paq{(1+4\xi)\de{1}+2\xi(\de{2}-1)}} \ , 
\ee
and, by further differentiating, the $\ep{i}$'s with arbitrary large $i$ can be obtained. In the large-$a$ limit, we have already calculated $\ep{2 i}=-n$ and $\ep{2i+1}=0$ [cf. Eqs. (\ref{ep11}) and (\ref{ep21})], and one then obtains two possible hierarchies for the $\de{i}$'s: 
\be{hi1}
\de{2i+1,\infty}=0,\;\de{2i ,\infty}=\ep{2,\infty}=-n \ , 
\ee
and
\be{hi2}
\de{1,\infty}=\frac{2\xi}{1+4\xi},\;\de{2i+1,\infty}=-n,\;\de{2i,\infty}=0 \ .
\ee
This latter statement cannot be simply verified by substitution because the limits involved do not commute. For example, on substituting first $\de{2}=0$ in Eq. (\ref{ep2IG}), one obtains $\ep{2}=0$, which is not correct. One must solve (at least perturbatively in the large-$a$ limit) Eq. (\ref{rel1IG}) and then evaluate the limits with the help of the solution found. The above results are correctly reproduced only if we proceed in this manner.

The exact reconstruction of the inflaton potential is not possible in general. Nonetheless, in specific cases, the potential may be derived exactly as follows. Consider the following ansatz for $\de{1}$:
\be{de1ans}
\de{1}=\frac{n_0+n_1a^{-n}}{d_0+d_1 a^{-n}} \ ,
\ee
which is suggested by the expression for $\ep{1}$ and Eq. (\ref{rel1IG}). If $n_0=0$ and $n_1\neq0$, then Eq. (\ref{de1ans}) can be integrated, and the resulting $\phi(a)$ is inverted as follows 
\be{inv1}
\phi(a)=\phi_0\pa{d_0+d_1 a^{-n}}^{-\frac{n_1}{n\,d_1}}\Rightarrow a^{-n}= \frac{\pa{\frac{\phi(a)}{\phi_0}}^{-\frac{n\,d_1}{n_1}}-d_0}{d_1} \ .
\ee
The coefficients $n_1$, $d_0$, $d_1$ and $\xi$ can finally be fixed by the requirement that Eq. (\ref{de1ans}) be a solution of Eq. (\ref{rel1IG}). Two nontrivial solutions can be found:
\be{s11}
d_0=-\frac{\pa{1+n}n_1\alpha}{A\,m\,n},\;d_1=-\frac{(1+n)n_1}{m\,n},\;\xi=\frac{m}{2(1-3m+n+m\,n)} \ ,
\ee
or
\be{s12}
d_0=-\frac{\pa{1+n}n_1\alpha}{A\,m\,n},\;d_1=-\frac{(1+n+m\,n)n_1}{m\,n},\;\xi=\frac{m}{2(1-2m+n+m\,n)} \ .
\ee
Notice that more exact solutions for $\de{1}$ can be found if we start from the ansatz (\ref{de1ans}) and $n_0\neq 0$. However, by further integrating these solutions to obtain $\phi(a)$, one is led to noninvertible functions, and the reconstruction cannot be completed. For both Eqs. (\ref{s11}) and (\ref{s12}) one has
\be{d1alarge2}
\de{1,\infty}=0 \ , 
\ee
and one can explicitly verify that the hierarchies belong to the set (\ref{hi1}). Notice that $n_1$ in Eqs. (\ref{s11}) and (\ref{s12}) can be arbitrarily chosen, as should be due to the form of the ansatz (\ref{de1ans}). Let us, for simplicity, complete the reconstruction choosing $n$ and $m$ to reproduce USR in the IG context ($n=6$, $m=1/2$). In this case, Eqs. (\ref{s11}) and (\ref{s12}) take the following form:
\be{s11usr}
n_0=0,\;d_0=-\frac{7\alpha}{3A}n_1,\;d_1=-\frac{7}{3}n_1,\;\xi=1/10\Rightarrow \de{1}=-\frac{3A}{7\pa{A+\alpha a^6}} \ ,
\ee
\be{s12usr}
n_0=0,\;d_0=-\frac{7\alpha}{3A}n_1,\;d_1=-\frac{10}{3}n_1,\;\xi=1/36\Rightarrow \de{1}=-\frac{3A}{10A+7\alpha a^6} .
\ee
From Eq. (\ref{Hans1}), $a(\phi)$ in (\ref{inv1}) and Eq. (\ref{s11usr}), one finds
\be{s11rec}
\de{1}=\frac{3}{7}\paq{\alpha\pa{\frac{\phi_0}{\phi}}^{14}-1}\quad {\rm and}\quad H^2=H_0^2\pa{\frac{\phi}{\phi_0}}^{14} \ , 
\ee
with $\phi/\phi_0\stackrel{a\rightarrow \infty}{\longrightarrow}\alpha^{1/14}$ and $\phi>\phi_0$, while for Eq. (\ref{s12usr}) one finds
\be{s12rec}
\de{1}=\frac{3}{10}\paq{\alpha\pa{\frac{\phi_0}{\phi}}^{20}-1}\quad {\rm and}\quad H^2=H_0^2\paq{\pa{\frac{7}{10}\frac{\phi}{\phi_0}}^{20}+\frac{3\alpha}{10}} \ , 
\ee
with $\phi/\phi_0\stackrel{a\rightarrow \infty}{\longrightarrow}\alpha^{1/20}$ and $\phi>\phi_0$. Finally, by using Eq. (\ref{VrecIG}), one obtains
\be{Vsol1}
V=-\frac{3H_0^2}{490\phi^{12}\phi_0^{14}}\pa{8\phi^{28}-72\alpha\phi_0^{14}\phi^{14}+15\alpha^2\phi_0^{28}}
\ee
for the first exact solution, and 
\be{Vsol2}
V=-\frac{H_0^2}{6000\phi^{38}\phi_0^{20}}\pa{7\phi^{20}+3\alpha\phi_0^{20}}\pa{7\phi^{40}-84\alpha\phi_0^{20}\phi^{20}+27\alpha^2\phi_0^{40}}
\ee
for the second. In the $a\rightarrow \infty$ limit, the potentials (\ref{Vsol1}) and (\ref{Vsol2}) satisfy the condition $\rd \ln V/\rd \ln \phi=4$. The potential can have negative values but, in the vicinity of $\phi\simeq \phi_\infty$, the potential is positive and $V_\infty>0$.

We discuss at last the behavior of the primordial scalar curvature spectrum. The general formulas illustrated in Sec. II can easily be generalised to the IG case wherein
\be{zIG}
z_{\rm IG}=a\phi\de{1}\sqrt{\frac{1+6\xi}{1+\de{1}}} \ ,
\ee
and $\Phi$ is given by
\be{PhiIG}
\Phi=\paq{1-\ep{1}-\frac{\ep{1}\ep{2}}{\pa{1-\ep{1}}}+\pa{2+2\de{1}+2\de{2}-\frac{\de{1}\de{2}}{1+\de{1}}}} \ .
\ee
If we evaluate $\Phi$ w.r.t. to the hierarchies (\ref{hi1}) and (\ref{hi2}), one observes that only constants and terms linear in the SR parameters remain. Moreover, $\ep{1,\infty}=0$ and $\Phi$ then simplifies to
\be{PhiIGsim}
\Phi=3+2\de{1}+2\de{2} \ , 
\ee
which can be negative only for the hierarchy (\ref{hi1}) but is strictly positive for the hierarchy (\ref{hi2}), provided we restrict ourselves to positive values of the nonminimal coupling $\xi$. In the former case, $\Phi=3-2n$, which implies that the growing solution exists for $n>3/2$.

If no growing solution exists [as is the case for (\ref{hi2}) or (\ref{hi1}) with $n<3/2$], an amplification of the spectrum is possible only if the spectrum is blue-tilted. Let us then evaluate $n_s-1$. In the IG case, $f_{\rm MS}(\ep{i})$ in the MS equation is given by
\be{fIG}
f_{\rm MS}=\de{1}^2+\de{2}^2+\pa{3-\ep{1}}\pa{1+\de{1}+\de{2}}+\de{2}\de{3}+\frac{\de{1}\de{2}\pa{\ep{1}+\de{1}-3\de{2}-\de{3}+\frac{2\de{1}\de{2}}{1+\de{1}}-2}}{1+\de{1}}-1 \ , 
\ee
and, as usual, it can be simplified to obtain the following expression for the scalar spectral index:
\be{nsm1IG}
n_s-1=3-\sqrt{1+4\pa{\de{1}^2+\de{2}^2+3\pa{1+\de{1}+\de{2}}-1}} \ .
\ee
Then, for the hierarchy (\ref{hi1}) and $n<3/2$, we obtain
\be{nsm1IGh1}
n_s-1=3-|3-2n|=2n \ , 
\ee
which is indeed blue-tilted, while for the hierarchy (\ref{hi2}), we find
\be{nsm1IGh2}
n_s-1=-\frac{4\xi}{1+4\xi} \ ,
\ee
which is red-tilted.

We therefore conclude that solutions having $H$ of the form given in Eq. (\ref{Hans1}), in the IG context, may lead to a spectrum enhancement for evolutions asymptotically described by the hierarchy (\ref{hi1}) and either for $n>3/2$ (due to the presence of the growing solution) or $0<n<3/2$ (in the absence of the growing solution but with the blue-tilted spectrum).

\section{Applications}
We have so far studied the consequences of cosmological evolutions with a transient phase, which is crucial to potentially obtain the amplification required by the formation of PBHs. Indeed, the presence of the transient generates, in the large-$a$ limit, a sequence of values for the SR parameters that is otherwise not obtained. We then reconstructed, when possible, the potentials that led to the desired evolution. In this section, our approach will be slightly different, as we shall study the presence of the transient solutions in the particular dynamical regime of constant roll (CR) inflation \cite{CR}, which is the natural generalization of USR.

CR solutions satisfy the equation
\be{CR}
\ddot \phi+B H\dot \phi=0 \ , 
\ee
where $B>0$, and one recovers the USR solution for $B=3$, while the case of $|B|\ll 1$ reproduces standard SR. We observe that the CR condition (\ref{CR}) can be rewritten, in terms of the SR parameters, as
\be{CRsr}
\de{2}+\de{1}-\ep{1}+B=0 \ .
\ee
Equation (\ref{CRsr}) is model independent, since it only depends on the definitions of $\ep{i}$'s and $\de{i}$'s, and can easily be integrated to obtain
\be{intCR}
\frac{\rd \phi}{\rd \ln a}H \pa{\frac{a}{a_0}}^B=C_3\Rightarrow \dot \phi=C_3\pa{\frac{a_0}{a}}^B \ , 
\ee
where $C_3$ is an integration constant.

In the minimally coupled case, CR can generate an amplification of the primordial scalar spectrum. In what follows, after a revision of this result (which was analyzed in \cite{GR}) we shall consider CR in IG and study its consequences.  

\subsection{Constant roll in GR with a minimally coupled inflaton}
In GR with a minimally coupled inflaton, on imposing CR conditions and adopting the Hamilton-Jacobi (HJ) formalism, it is possible to reconstruct the evolution of the Hubble parameter and the corresponding potential \cite{GR}. In particular, one finds that $H(\phi)$ is the following superposition of two exponential functions:
\be{Hphigr}
H(\phi)=C_1\exp\pa{\sqrt{\frac{B}{2}}\frac{\phi}{\M}}+C_2\exp\pa{-\sqrt{\frac{B}{2}}\frac{\phi}{\M}} \ .
\ee
In \cite{GR}, the solution (\ref{Hphigr}) with one exponential ($C_1=0$ or $C_2=0$), as well as the $\cosh$ and $\sinh$ cases, are analyzed with the aim of finding the exact solutions compatible with CMB observations \cite{Planck} (and thus not amplified). In particular, Ref. \cite{CR2} focuses on the $\cosh$ case, which corresponds to the $C_1=C_2\neq 0$ case, and implements it for a potential that exhibits two stages of slow roll that are separated by a constant roll phase, within a framework that also fulfills the current power-spectra constraints.

Here, in a slightly different approach, we consider the general case, and we study the power enhancement of the spectrum. Equation (\ref{fr1}) can be rewritten in terms of the SR parameters
\be{fr1eq}
\ep{1}=\frac{\phi^2}{2\M^2}\de{1}^2 \ , 
\ee
from which, using the chain rule, we obtain
\be{ep1H}
\ep{1}=-\de{1}\frac{\rd \ln H}{\rd \ln \phi} \ . 
\ee
Equation (\ref{fr1eq}) then becomes
\be{fr1eq2}
\ep{1}=\frac{2\M^2}{\phi^2}\pa{\frac{\rd \ln H}{\rd \ln \phi}} \ .
\ee
The potential can subsequently be reconstructed by substituting Eqs. (\ref{Hphigr}) and (\ref{fr1eq2}) into Eq. (\ref{fr01}):
\be{VgrCR}
V(\phi)=\M^2H(\phi)^2\paq{3-\frac{2\M^2}{\phi^2}\pa{\frac{\rd \ln H}{\rd \ln \phi}}^2} \ .
\ee 
To obtain the corresponding evolution, one must integrate and invert the equation 
\be{eqGRCRtoinv}
\de{1}=-\frac{2\M^2}{\phi^2}\frac{\rd \ln H}{\rd \ln \phi} \ , 
\ee
which can easily be derived from Eq. (\ref{fr1eq}) by using (\ref{ep1H}). One finds
\be{aphiCR}
\pa{\frac{a}{a_0}}^B=\frac{x}{B\pa{C_2-C_1 x^2}} \ , 
\ee
where $x=\exp\pa{\sqrt{\frac{B}{2}}\frac{\phi}{\M}}$. It is straightforward to invert Eq. (\ref{aphiCR}) so as to obtain $x=x(a)$. Correspondingly, one has 
\be{HaCR}
H(a)=\pm\frac{4C_1C_2+\pa{\frac{a_0}{a}}^{2B}\mp \pa{\frac{a_0}{a}}^{B} \sqrt{4 C_1C_2+\pa{\frac{a_0}{a}}^{2B}}}{\mp \pa{\frac{a_0}{a}}^{B}+\sqrt{4 C_1C_2+\pa{\frac{a_0}{a}}^{2B}}}\stackrel{a\rightarrow \infty}{\longrightarrow} \pm\frac{8C_1C_2+\pa{\frac{a_0}{a}}^{2B}}{4\sqrt{C_1C_2}} \ .
\ee
Notice that the same result can be obtained if one uses the CR definition (\ref{CRsr}) instead of Eq. (\ref{fr1}).

The last, approximate, equality in Eq. (\ref{HaCR}) is the large-$a$ limit of $H(a)$, and this shows that the CR evolution is asymptotically equivalent to the evolution given in Eq. (\ref{Hans1}) with $m=1$ and $n=2B$. The results obtained in Sec. II for large $a$ are therefore inherited by CR. Thus, one obtains $\ep{2i+1,\infty}=0$ and $\ep{2i,\infty}=2B$. Correspondingly, 
\be{PhiCR}
\Phi=3-2B \ ,
\ee
which shows that the curvature perturbations are amplified for $B>3/2$ due to the presence of a growing solution. In contrast, if $0<B<3/2$, one finds a blue-tilted spectrum
\be{nsm1CR}
n_s-1=3-\sqrt{(3-2B)^2}=2B>0 \ , 
\ee
i.e., a spectrum enhancement in the absence of growing solutions. Therefore, CR inflation admits transient solutions that always lead to an amplification. Finally, it is worthwhile to mention that the solutions with $C_1=0$ or $C_2=0$ simply correspond to the attractor solutions for power law inflation, and thus they are not associated with any amplification effect. Indeed, in Ref. \cite{CR2}, a $\cosh$ type potential and the corresponding transient solution is considered in order to obtain the desired enhancement of the primordial spectra in the CR scenario.

\subsection{Constant roll with a nonminimally coupled inflaton}
Let us now consider CR in the IG context. For this case, the HJ formalism leads to \cite{ST}
\be{HCRIG}
H(\phi)=C_1\phi^{(B+p)/2}+\frac{C_2}{\phi^{(p-B)/2}} \ , 
\ee
where $p=\sqrt{(B+2)^2+2B(2+\xi^{-1})}$ and $(B+p)/2$ and $(p-B)/2$ are both positive with $(p-B)/2<(B+p)/2$.  For simplicity, we shall take $C_{1,2}>0$ and we restrict the analysis to the $\phi>0$ interval. Studying the spectrum enhancement for CR in the IG case is more complicated than for GR. This is essentially a consequence of the complicated form of Eq. (\ref{rel1IG}) in comparison to Eq. (\ref{fr1}) in the GR case. However, the simple relation (\ref{CRsr}) holds, and it can be used to simplify the equations. First, with Eq. (\ref{CRsr}), one may eliminate $\de{2}$ from Eq. (\ref{rel1IG}) and obtain
\be{rel1CR}
\ep{1}=\frac{1+2\xi}{2\xi}\de{1}^2-(B+1)\de{1} \ . 
\ee
Subsequently, by using Eq. (\ref{ep1H}), one finds 
\be{d1CR}
\de{1}=\frac{2\xi}{1+2\xi}\pa{B+1-\frac{\rd \ln H}{\rd \ln \phi}} \ ,
\ee
and the potential can be reconstructed by substituting Eqs. (\ref{HCRIG}) and (\ref{d1CR}) into Eq. (\ref{VrecIG}).

The evolution could be obtained by integrating Eq. (\ref{d1CR}) and inverting the result. However, analytically inverting the resulting equation for arbitrary values of the parameters is impossible. As we are only interested in the asymptotic form of $H(a)$, one can employ a perturbative approach. Integration of Eq. (\ref{d1CR}) yields
\be{aphiIG}
\pa{\frac{a_0}{a}}^B=\phi^{\frac{2+B}{2}}\paq{\pa{B+p+2}C_1\phi^{\frac{p}{2}}+\pa{B-p+2}\frac{C_2}{\phi^{\frac{p}{2}}}} \ , 
\ee
where $B+p+2>0$ and $B-p+2<0$. Therefore, in the large-$a$ limit, the inversion of Eq. (\ref{aphiIG}) leads to
\be{phiapert}
\phi(a)=\phi_\infty+\sum_{i>1}\phi_i \pa{\frac{a_0}{a}}^{i\,B}\sim \phi_\infty+\phi_1 \pa{\frac{a_0}{a}}^{B} \ ,
\ee
where $\phi_\infty$ is positive. By substituting Eq. (\ref{phiapert}) into Eq. (\ref{HCRIG}) and expanding for large $a$ (properly accounting for the next-to-leading-order contributions), one finally obtains the asymptotic form of $H(a)$, which reads
\be{HinfIG}
H\sim H_\infty+H_1 \pa{\frac{a_0}{a}}^{B} \ .
\ee
Comparison to Eq. (\ref{Hans1}) shows that $m=1$ and $n=B$, and the corresponding hierarchy of $\de{i}$'s is given by Eq.(\ref{hi1}) since  
\be{asyde1}
\de{1,\infty}=\frac{\lim_{a\rightarrow \infty}\dot \phi}{H_{\infty}\phi_\infty}=0 \ ,
\ee
where $\dot \phi$ is given by (\ref{intCR}). One then obtains 
\be{CRIGinf1}
\Phi=3-2B, \;n_s-1=2B \ ,
\ee
and, when $0<B<3/2$,
\be{CRIGinf2}
n_s-1=2B \ ,
\ee
which are the same results as GR with a minimally coupled inflaton. Indeed, in the $a\rightarrow \infty$ limit, the homogeneous inflaton is frozen at a certain value and one essentially recovers the evolution of the minimally coupled case, where ``Newton's constant'' is now reproduced by the (constant) asymptotic value of the inflaton. Furthermore the $a$ dependence of the solution is a consequence of the fact that the CR condition (\ref{CRsr}) is independent of the specific inflationary model, provided $H_\infty$ and $\phi_\infty$ are found to be (finite) constants.

\subsection{Jordan and Einstein frame mapping}
In the previous section, we found the same asymptotic behavior for the spectra in the minimally coupled case and in the IG case. This result was obtained in spite of the fact that the CR condition is not frame invariant; i.e., the CR condition in the Einstein frame (EF) is not mapped, in general, into a CR condition in the Jordan frame (JF). In this section, we briefly review this statement and discuss its consequences.

It is well known that, by a suitable conformal transformation and a redefinition of the scalar field (inflaton), one can map a minimally coupled theory (defined in the so-called EF) into a nonminimally coupled one (in the JF). In particular, for IG, the mapping is given by the following transformation rules (see, e.g., \cite{V}):
\be{metricEJ}
a(t)=\frac{\M}{\sqrt{\xi}\sigma}\tilde a(t),\; N(t)=\frac{\M}{\sqrt{\xi}\sigma}\tilde N(t) \ , 
\ee 
and 
\be{EJ}
\phi=\M\sqrt{\frac{1+6\xi}{\xi}}\ln\frac{\sigma}{\sigma_0}, \tilde V(\phi(\sigma))=\frac{\M^2}{\xi^2\sigma^4} V(\sigma) \ ,
\ee
where the tilde refers to the Einstein frame, $\phi$ is the scalar field in the EF, and $\sigma$ is that in the JF. Notice that here $N(t)$ and $\tilde{N}(t)$ in Eq. (\ref{metricEJ}) correspond to the lapse function in the Jordan and Einstein frames, respectively, and they are not to be confused with the number of $e$-folds, which was previously denoted by $N$. The mapping induces the following transformations of the Hubble parameter:
\be{hEF}
\tilde H=\frac{\rd \tilde a/\rd t}{\tilde N\tilde a}=\pa{1+\de{1}}\frac{\M}{\sqrt{\xi}\sigma}H \ ,
\ee
where
\be{JFh1}
H(t)=\frac{\rd a(t)/\rd t}{a(t) N(t)},\; \ep{i+1}=\frac{\rd \ep{i}/\rd t}{\ep{i} N(t) H(t)},\; \de{i+1}=\frac{\rd \de{i}/\rd t}{\de{i} N(t) H(t)}
\ee
are the Hubble and SR parameters in the JF. From the relation (\ref{metricEJ}), one also finds that
\be{logaEJ}
\frac{\rd}{\rd \ln \tilde a}=\pa{1+\de{1}}^{-1}\frac{\rd}{\rd \ln a}\ .
\ee
It is now straightforward to obtain the relations between SR parameters in the two frames:
\be{ep1}
\tep{1}\equiv-\frac{\rd \ln \tilde H}{\rd \ln \tilde a}=-\pa{1+\de{1}}^{-1}\frac{\rd}{\rd \ln a}\ln \paq{\pa{1+\de{1}}\frac{\M}{\sqrt{\xi}\sigma}H}=\frac{\de{1}+\ep{1}-\frac{\de{1}\de{2}}{1+\de{1}}}{1+\de{1}} \ .
\ee
Given the relation (\ref{rel1IG}), one then finds
\be{relEJ}
\tep{1}=\frac{(1+6\xi)\de{1}^2}{2\xi(1+\de{1})^2} \ .
\ee
From Eqs. (\ref{rel1IG}) and (\ref{relEJ}), given that CR for a minimally coupled inflaton has $\tep{1,\infty}=0$, one concludes that, correspondingly, in the JF one has $\de{1,\infty}=0$ and $\ep{1,\infty}=0$. If we differentiate Eq. (\ref{relEJ}), we obtain the following relations among other SR parameters in the two frames
\be{tep2}
\tep{2}=\frac{2\de{2}}{(1+\de{1})^2} \ ,
\ee
\be{tep3}
\tep{3}=\frac{\de{3}-2\de{1}\de{2}+\de{1}\de{3}}{\pa{1+\de{1}}^2} \ ,
\ee
\be{tep4}
\tep{4}=\frac{\de{1}\de{2}\paq{2\de{2}-2\de{1}\de{2}+3\de{3}+3\de{1}\de{3}-\pa{1+\de{1}}^2\de{3}\de{4}}}{\pa{1+\de{1}}^2\pa{2\de{1}\de{2}-\de{1}\de{3}-\de{3}}} \ ,
\ee
and further $\tep{i}$'s can be found by iterating the procedure but are useless for what follows.

Similarly, one can directly calculate the relations of the $\tde{i}$'s with the dynamical variables in the JF:
\be{deftde1}
\tde{1}\equiv \frac{\dot\phi}{\tilde N \tilde H\phi}=\sqrt{\frac{1+6\xi}{\xi}}\frac{\M}{\phi}\frac{\de{1}}{1+\de{1}} 
\ee
and 
\be{deftde2}
\tde{2}\equiv\frac{\rd \tde{1}/\rd t}{{\tilde N}{\tilde H}\tde{1}}=-\tde{1}+\frac{\de{2}}{\pa{1+\de{1}}^2} \ .
\ee
From the last relation and Eq. (\ref{rel1IG}), one has that the CR condition in the EF,
\be{CREF}
\tde{2}+\tde{1}-\tep{1}+B=0 \ ,
\ee
is mapped into the following condition in the JF:
\be{CRmapJF}
\de{2}+\pa{B-1}\de{1}-\ep{1}+B=0 \ .
\ee
Notice that only for $B=2$ the CR condition is frame invariant. Nonetheless, both equations reduce to $\de{2,\infty}=\tde{2,\infty}=-B$ at late times, and the evolution is indistinguishable, at least as far as the homogeneous degrees of freedom and the inflationary spectra are concerned.

We conclude that, whereas the scalar spectral index $n_s-1$ is frame invariant, $\Phi$ is generally not frame invariant. This can be checked directly by substitution. However, assuming CR holds in the EF, one verifies that $\Phi$ and $n_s-1$ are both frame invariant in the asymptotic regime. This approximate invariance can be intuitively understood from the fact that, for large scale factors (late times), the CR condition in the EF implies that the nonminimally coupled homogeneous inflaton in JF freezes at a particular asymptotic value (due to $\de{1}\to0$), which can be seen as ``Newton's constant'', so as to recover the evolution of the minimally coupled EF scenario [see comments after Eq. (\ref{CRIGinf2})]. Indeed, one may verify by substitution that $\Phi$ is frame invariant in the $\de{1}\rightarrow 0$ limit (i.e., when the field in the Jordan frame freezes).
\begin{table}[h!]
  \begin{center}
    \caption{Results summary}
    \label{table1}
    \begin{tabular}{cccc} 
      \textbf{Inflation} & \textbf{Asymptotic} & \textbf{Growing} & \textbf{Blue-tilted}\\
      \textbf{model} & \textbf{solution} & \textbf{solution} &\textbf{spectral index}\\
      \hline
      GR& dS & $n>3$ & $0<n<3$\\
      GR & PL & $-$ & $-$\\
      IG & dS, $\de{1,\infty}=0$ & $n>3/2$ & $0<n<3/2$\\
      IG & dS, $\de{1,\infty}\neq 0$ & $-$ & $-$\\
      CR+GR & dS & $B>3/2$ & $0<B<3/2$\\
      CR+IG & dS, $\de{1,\infty}=0$ & $B>3/2$ & $0<B<3/2$\\
    \end{tabular}
  \end{center}
\end{table}

\section{Conclusions}
In this article, we have analyzed the effects of different transient phases, which may occur during inflation due to a particularity of the inflaton potential, on the primordial inflationary spectrum of scalar perturbations. These transients have been studied in the last few years as sources of amplification of the amplitude of the curvature spectrum. It is important to notice that if the amplitude of scalar perturbations grows large enough, it may induce gravitational collapse and consequently seed the formation of primordial black holes after inflation ends. In the literature, several mechanisms for such an amplification during inflation have been proposed. In particular, the presence of an ultraslow-roll or, more generally, a constant-roll phase has been studied. Whereas in the former case the amplification is due to the existence of a growing solution to the equation of motion of the curvature perturbations, in the latter case the amplification can also be generated by a blue-tilted spectrum in the absence of the growing solution.

The purpose of this paper was precisely to examine general features of the aforementioned models starting from a rather generic ansatz for the Hubble parameter as a function of the scale factor. This general description of the transient phase is model independent, and many results obtained can readily be applied to several modified gravity models. The matter-gravity dynamics is described in terms of the hierarchies of SR parameters, both at the homogeneous level and at the level of perturbations. These hierarchies, when the transient phase that describes the approach to some inflationary attractor is considered, have been shown to take a peculiar form wherein either odd or even terms of the hierarchy are null and the remaining ones are different for zero. This general feature is a peculiarity of the asymptotic form of the SR parameters close to the attractor, and it is then used as a simplifying assumption throughout the entire article. The resulting hierarchies, in the large-$a$ limit and for the cases considered, were used to calculate the behavior of the primordial curvature spectrum as the parametrization of $H(a)$ was varied. Then, when possible, the corresponding inflaton potential was fully reconstructed. An overview of the spectra enhancement results was presented in Table \ref{table1}.

For simplicity, only the induced gravity case has been considered here as a generalization of general relativity with a minimally coupled inflaton. Induced gravity is particularly relevant since both Higgs and Starobinsky inflationary models (which are in good agreement with observations) take place in the `induced gravity phase'. We note that while transient evolutions that have the de Sitter universe as a limit (such as USR) can lead to an amplification, the results differ when power law inflation is considered as the limit of a transitory dynamics and, for the cases we were able to solve explicitly, no modification of the scalar spectrum was obtained. Finally, the constant-roll case was discussed in more detail as an application of the preceding results, and the issue of the transition from the Einstein frame to the Jordan frame was also scrutinized.

\vspace{-0.5cm}
\begin{acknowledgements}
L.C. thanks the Dipartimento di Fisica e Astronomia of the Universit\`{a} di Bologna as well as the I.N.F.N. Sezione di Bologna for financial support.
\end{acknowledgements}
\end{document}